\documentstyle[12pt,sw20lart]{article}
\begin{document}

\title{{\Large {\bf Ricci and Matter inheritance collineations of Robertson-Walker
space-times}}}
\author{Pantelis S. Apostolopoulos\thanks{%
E-mail: papost@cc.uoa.gr} \quad and Michael Tsamparlis\thanks{%
E-mail: mtsampa@phys.uoa.gr} \and {\it \ {\small Department of Physics,
Section of Astrophysics-Astronomy-Mechanics, }} \and {\it {\small University
of Athens, Zografos 15783, Athens, Greece} }}
\maketitle

\begin{abstract}
It is well known that every Killing vector is a Ricci and Matter
collineation. Therefore the metric, the Ricci tensor and the energy-momentum
tensor are all members of a large family of second order symmetric tensors
which are invariant under a common group of symmetries. This family is
described by a generic metric which is defined from the symmetry group of
the space-time metric. The proper Ricci and Matter (inheritance)
collineations are the (conformal) Killing vectors of the generic metric
which are not (conformal) Killing vectors of the space-time metric. Using
this observation we compute the Ricci and Matter inheritance collineations
of the Robertson-Walker space-times and we determine the Ricci and Matter
collineations without any further calculations. It is shown that these
higher order symmetries can be used as supplementary conditions to produce
an equation of state which is compatible with the Geometry and the Physics
of the Robertson-Walker space-times.

{\bf Date of submission: 02-10-2001}
\end{abstract}

\section{Introduction}

A geometric symmetry or collineation is defined by a relation of the form:

\begin{equation}
{\cal L}_{{\bf \xi }}{\bf \Phi }={\bf \Lambda }  \label{sx1.1}
\end{equation}
where $\xi ^{a}$ is the symmetry or collineation vector, ${\bf \Phi }$ is
any of the quantities $g_{ab},\Gamma _{bc}^{a},R_{ab},R_{{}\mbox{ }bcd}^{a}$
and geometric objects constructed from them and ${\bf \Lambda }$ is a tensor
with the same index symmetries as ${\bf \Phi }$. By demanding specific forms
for the quantities ${\bf \Phi }$ and ${\bf \Lambda }$ one finds all the well
known collineations. For example $\Phi _{ab}=g_{ab}$ and $\Lambda
_{ab}=2\psi g_{ab}$ defines the Conformal Killing vectors (CKV) and
specializes to a Special Conformal Killing vector (SCKV) when $\psi _{;ab}=0$%
, to a Homothetic vector field (HVF) when $\psi =$constant and to a Killing
vector (KV) when $\psi =0.$ When $\Phi _{ab}=R_{ab}$ and $\Lambda
_{ab}=2\psi R_{ab}$ the symmetry vector $\xi ^{a}$ is called a Ricci
Inheritance Collineation (RIC) and specializes to a Ricci Collineation (RC)
when $\Lambda _{ab}=0$. When $\Phi _{ab}=T_{ab}$ and $\Lambda _{ab}=2\psi
T_{ab},$ where $T_{ab}$ is the energy momentum tensor, the vector $\xi ^{a}$
is called a Matter Inheritance Collineation (MIC) and specializes to a
Matter collineation (MC) when $\Lambda _{ab}=0$. The function $\psi $ in the
case of CKVs is called the conformal factor and in the case of inheriting
collineations the {\em inheriting factor}.

There are other important families of collineations and in fact most (but
not all) of them have been classified in a tree like diagram which exhibits
clearly their relative properness \cite{Katzin-Levine-Davis,Katzin-Levine}.
Collineations of a different type are not independent, for example a KV\ is
a RC\ or MC but not the opposite. Using this simple observation we approach
collineations form a different perspective which groups seemingly unrelated
collineations in one, thus changing the above classification and in fact 
{\em collapsing the classification tree} considerably. The core of our
argument lies in the definition of a metric by its symmetry algebra.

Assume that a manifold ${\cal M}$ admits the symmetry (Lie) group $G$ of
dimension $r$ whose associated Lie algebra is ${\cal G}$. Then there exists
on ${\cal M}$ a symmetric non-degenerate second order tensor (which we call 
{\em generic metric}) invariant under the group $G$. The action of the
symmetry group $G$ is characterized by the dimension $d$ of the group
orbits. If $s$ is the dimension of the stability (isotropy) group then \cite
{Eisenhart} $r=d+s$. In the above scenario the type of the orbits (that is
the signature of the metric on the orbits) does not appear explicitly and it
is fixed, in general, by an extra assumption. This analysis leads one to
consider the following ''inverse'' problem.

Suppose one is given a symmetry group of a space-time metric (that is a
group of KVs) of dimension $r$ whose orbits have everywhere dimension $d$.
Let $\{{\bf X}_{1},...,{\bf X}_{r}\}$ be the KVs generating the Lie algebra
of $G$. Find all non degenerate, second order symmetric tensors, $G_{ab}$
say, which satisfy the equations ${\cal L}_{{\bf X}_{A}}G_{ab}=0$ where $%
A=1,2,...,r$, that is, their symmetry algebra is isomorphic to ${\cal G}$.

Obviously the metric of the space-time manifold belongs to the family of
these tensors but, as we shall see below, there are more. In order to solve
the problem one observes that the data \{Symmetry algebra ${\cal G}$,
dimension of orbits\} define a generic metric (=second order tensor) which
can be constructed as follows:

a. The ''part'' of the metric on the group orbits has the same functional
form as the original space-time metric but not a specific signature.

b. Outside the group orbits, only the functional dependence of the
components of the metric will be restricted, that is, they will be
independent of the coordinates defined on the orbits i.e. if one selects
coordinates adapted to the KVs of the metric then the components of the
required tensors in the quotient submanifold ${\cal M}/orbit$ do not depend
on these coordinates.

The above construction implies that all symmetric tensors we are looking
for, are obtained from a generic element provided that the orbits are {\em %
non null}. In the following the set of all second order symmetric tensors
which admit the symmetry algebra of a space-time metric as their symmetry
algebra we shall call {\em family of metrics} {\em (FOM)} and the generic
element describing this set {\em generic metric.} The later we shall denote $%
G_{ab}$.

One can break the FOM into disjoint subsets by demanding specific signatures
on the orbit part of the generic metric. The KVs of the elements in each
subset are, in general, functionally different but all belong to the
symmetry algebra of the FOM. This implies that in order to compute the
complete algebra of KVs of the elements in each subset, it is enough to
solve Killing's equations for the FOM and then select the KVs for various
values of the signature. In this case the complete Lie algebra of KVs
contains as a subalgebra the symmetry algebra ${\cal G}$ of the original
space-time metric.

The above considerations can be useful in two important cases.

\underline{The Ricci Tensor}

Suppose one has a space-time (the argument is independent of the signature
and holds for non-space-time metrics as well) metric $g_{ab}$ which admits a
certain Lie algebra ${\cal G}$ of KVs. Let $G_{ab}$ the generic metric
defined by the symmetry algebra ${\cal G}$. The associated Ricci tensor $%
R_{ab}$ of the metric $g_{ab}$ is computed by covariant differentiation
(geometry only!) and is a second order symmetric tensor on the space-time
manifold. It is well known that the KVs of a metric are RCs, therefore the
Ricci tensor inherits the symmetries of the metric. This implies that {\em %
the Ricci tensor and the original space-time metric g}$_{ab}$ {\em belong to
the FOM generated by the }$G_{ab}$. Furthermore the Ricci tensor must follow
from the generic metric $G_{ab}$ for some values of its defining parameters
and the signature. In conclusion:

a. The form of the Ricci tensor is restricted by the symmetry (must belong
to one of the subsets of the FOM defined by $G_{ab}$).

b. The symmetries of the Ricci tensor are computed from the KVs of $G_{ab}$.
The KVs of the generic metric which are not, in general, KVs of the
space-time metric $g_{ab}$ are precisely the proper RCs. In the following we
shall call these vectors {\em extra} KVs of the generic metric $G_{ab}$.
Hence in order to compute the proper RCs of a given space-time metric $%
g_{ab} $ it is enough one to compute the {\em extra} KVs of the generic
metric $G_{ab}$ and subsequently express them in terms of the components of $%
R_{ab}$ (replacing those of the generic metric).

In Section V we shall apply the above conclusions and we shall compute all
proper RCs of the RW space-times in a straightforward and simple manner.

\underline{The Energy Momentum Tensor}

Let us consider a space-time metric $g_{ab}$ which is created by a matter
distribution described by the energy-momentum tensor $T_{ab}$. From Einstein
field equations (Geometry combined with Physics!) we have that the
symmetries of the metric pass over to the energy-momentum tensor. Therefore
what has been said for the Ricci tensor also applies to the energy momentum
tensor. This has, among others, the following important implications:

a. In a given space-time admitting a group of KVs, the energy momentum
tensor is not arbitrary but {\em it must} be an element in the FOM defined
by that symmetry group. This implies that the symmetry assumptions do not
fix only the geometry of space-time but also its dynamics, an interplay much
desired in General Relativity.

b. In order to compute the proper MCs of a given space-time one simply
expresses the components of the extra KVs of the generic metric $G_{ab}$ in
terms of the components of $T_{ab}.$

It must be emphasized that the Ricci tensor and the energy-momentum tensor
although formally the same from the point of view of symmetry are different
in two important aspects. The first is the energy conditions which apply
directly to the energy-momentum tensor and indirectly on the Ricci tensor
via the Einstein field equations. The second is the physical interpretation
of the energy-momentum tensor in terms of dynamic variables (energy density,
pressure etc.) which is achieved by the introduction of the fluid four
velocity $u^{a}$ ($u^{a}u_{a}=-1$) and the subsequent well known 1+3
decomposition \cite{Ellis1}:

\begin{equation}
T_{ab}=\mu u_{a}u_{b}+ph_{ab}+2q_{(a}u_{b)}+\pi _{ab}  \label{sx1.2}
\end{equation}
where $h_{ab}=g_{ab}+u_{a}u_{b}$ is the projection tensor. We shall use
these observations later when we discuss the application of matter
collineations in RW space-times.

What we have said about the KVs of the space-time metric extends without any
change to CKVs which also form a group of symmetries (the conformal symmetry
group). This means again that one can compute the proper RICs or MICs from
the extra CKVs of the generic metric, that is, the CKVs which are not CKVs
of the space-time metric.

In order to demonstrate the validity and the importance of the above
considerations and, at the same time, produce useful practical results, in
subsequent sections we work with the Robertson-Walker (RW) space-times.
Indeed, besides the fact that the RW space-time metric is used as the
standard model for cosmological observations, very few of the RCs and the
MCs it admits are known. Green et al. \cite{Green-Norris-Oliver-Davis} and
more recently Nunez et al. \cite
{Nunez-Percoco-Villalba,Conteras-Nunez-Percoco} found some RCs. As far as we
aware there do not exist results concerning the MCs in RW space-times and
only recently Carot et al. \cite{Carot-daCosta-Vaz} have studied MCs in a
general context. Inherited collineations other than CKVs are relatively new
and have not been considered much in the literature. Most of the existing
work has been done by Duggal et al. \cite
{Duggal1,Duggal2,Sharma-Duggal,Duggal3} Because inherited collineations are
more general than simple collineations one expects that perhaps they could
possibly be useful in various ways.

One might ask the further question: Why should we be interested in these
collineations? What could be their use? Let us answer this question in the
case of the standard cosmological model. Using the symmetry assumptions of
the RW metric and Einstein Field Equations (EFE) one ends up with a
differential equation involving the scale factor and one of the two dynamic
quantities $\mu ,p.$ This equation cannot be solved unless one introduces an
equation of state $p=p(\mu )$. The first (and logical choice) is a linear
equation of state to which one gives physical meaning by considering rather
`ideal'' matter forms. It would be preferable for one to use non-linear
equations of state, which could express more complex matter situations. But
how would these equations be constructed?

The above results suggest such a method: Use a geometric condition to get
the desired equation and then study its physical significance posteriori. As
we have shown the symmetries couple the geometry ($R_{ab}$) and the matter ($%
T_{ab}$) to the gravitational field ($g_{ab}$) in an inherent manner.
Therefore one can use these ''higher'' symmetries and let the geometry
produce the Physics, if any. In other words one requires that the space-time
admits (e.g.) a RC and produces from that demand an equation of state which
leads to a solution of the equation defining the scale factor, therefore to
a definite cosmological model.

An outline of the paper is as follows: In section II we determine the
conformal algebra of a general space which is conformally related to a $%
1+(n-1)$ decomposable space and whose $(n-1)$\ subspaces are spaces of
constant curvature $(n>2)$. In section III we apply the general results to
the case $n=4$. Demanding the signature to be -$2$ we obtain the conformal
algebra of the RW\ space-time regaining in a natural manner known results 
\cite{Maartens-Maharaj}. For the case where the signature of the $1+3$
metric is $4$ we determine the conformal algebra of the resulting
4-dimensional Euclidean space. The main results of the paper are in Sections
IV, V, and VI which give the {\em complete} algebra of RICs of the RW
space-times both in the degenerate and in the non-degenerate cases (Section
IV) and the proper RCs, MICs and MCs (Sections V and VI respectively). In
order to get a physical sense of the results we consider, in Section VII,
the standard comoving observers and we study the evolution of the flat RW
space-times which admit MCs. Finally Section VIII\ concludes the paper.

\section{The conformal algebra of metrics conformal to a $1+(n-1)$
decomposable metric}

In the determination of the RICs and MICs of the RW space-times we shall
need the complete conformal algebra of a 4-dimensional metric which is
conformally related to a 1+3 decomposable metric whose 3-spaces are spaces
of constant curvature (and signature $3$).

The conformal algebra of the standard RW space-times has been derived before 
\cite{Maartens-Maharaj} using mainly the Lie algebra property of CKVs. In
this work we compute this algebra using a different and more general
approach. The presentation will be concise and most of the results will be
taken directly from the existing literature. We work in an $n-$dimensional
space and in the next section we restrict the results to the case $n=4$ .

Consider two metrics (in an $n-$dimensional space) which are conformally
related:

\begin{equation}
\widehat{g}_{ab}=U^{2}(x^{c})g_{ab}.  \label{sx2.1}
\end{equation}
It is well known \cite{Defrise-Carter,Hall-Steele} that these metrics have
the same contravariant CKVs $X^{a}$ and that the covariant vectors $\widehat{%
X}_{a}=\widehat{g}_{ab}X^{b},$ $X_{a}=g_{ab}X^{b}$ are related as follows $%
\widehat{X}_{a}=U^{2}(x^{c})X_{a}$. This implies for the conformal factor
and the bivector of these vector fields the relations: 
\begin{equation}
\widehat{\phi }=X^{a}(\ln U)_{,a}+\phi  \label{sx2.2}
\end{equation}

\begin{equation}
\widehat{F}_{ab}=U^{2}F_{ab}-2UU_{[,a}X_{b]}  \label{sx2.3}
\end{equation}
where we have used an obvious notation.

A CKV is a gradient CKV if the corresponding bivector vanishes. Let us
assume that the second metric is flat (of arbitrary sign). Then it is well
known \cite{Bruhat} that the CKVs are:

\begin{equation}
X_{c}=a_{c}+a_{cd}x^{d}+bx_{c}+2(b_{d}x^{d})x_{c}-b_{c}(x^{d}x_{d})
\label{sx2.4}
\end{equation}
where the constants $a_{c},a_{cd}=-a_{dc},b,b_{c}$ specify respectively the
type (i.e. KV, HVF, SCKV) of the CKV. If we further assume the metric $%
\widehat{g}_{ab}$ to be the metric of a space of constant curvature then $%
U(x^{c})=\left( 1+\frac{k}{4}x^{c}x_{c}\right) ^{-1}$ where $k=\frac{R}{%
n(n-1)}$, $R$ is the (constant)\ scalar curvature of the metric and $n$ is
the dimension of the space. In this case it can be shown \cite
{Tsamparlis-Nikolop-Apost} that the metric $\widehat{g}_{ab}$ admits $(n+1)$ 
{\em gradient} CKVs whose conformal factors satisfy the equation:

\begin{equation}
\widehat{\phi }_{;ab}=-k\widehat{\phi }\widehat{g}_{ab}.  \label{sx2.5}
\end{equation}
Using the above information one can derive easily the $(n+1)(n+2)/2$ CKVs
together with their bivector and conformal factor listed in Table I.
\pagebreak

{\small TABLE I. The }$(n+1)(n+2)/2${\small \ CKVs of an n-space of constant
non-vanishing curvature. }$\delta _{\alpha \beta }^{\kappa \lambda }\equiv
\delta _{[\alpha }^{\kappa }\delta _{\beta ]}^{\lambda }${\small \ and the
non-tensorial indices }$\alpha ,\beta ,\lambda ,\nu =1,...,n${\small \ count
vector fields.}

\begin{center}
$
\begin{tabular}{|c|c|c|c|c|}
\hline
$
\begin{array}{l}
\mbox{{\bf Number of}} \\ 
\mbox{{\bf CKVs}}
\end{array}
$ & {\bf CKVs of the }${\bf n}${\bf -metric} & $F_{\alpha \beta }$ & $\phi $
& $\phi _{,\alpha }$ \\ \hline
$n$ & ${\bf I}_{\nu }=\frac{1}{U}\left[ (2U-1)\delta _{\nu }^{\alpha }+\frac{%
1}{2}kUx_{\nu }x^{\alpha }\right] \partial _{\alpha }$ & $2kU^{3}x_{[\alpha
}\delta _{\beta ]\nu }$ & $0$ & $0$ \\ \hline
$n(n-1)/2$ & ${\bf M}_{\lambda \nu }=\delta _{\lambda \nu }^{\alpha \beta
}x_{\alpha }\partial _{\beta }$ & $U^{2}\delta _{\alpha \beta }^{\lambda \nu
}+kU^{3}x_{[\alpha }\delta _{\beta ]\sigma }^{\lambda \nu }x^{\sigma }$ & $0$
& $0$ \\ \hline
$1$ & ${\bf H}=x^{\alpha }\partial _{\alpha }$ & $0$ & $1-\frac{kU\cdot
(x^{\sigma }x_{\sigma })}{2}$ & $-kH_{\alpha }$ \\ \hline
$n$ & ${\bf C}_{\nu }=\frac{1}{U}\left[ \delta _{\nu }^{\alpha }-\frac{%
kUx_{\nu }x^{\alpha }}{2}\right] \partial _{\alpha }$ & $0$ & $-kUx_{\nu }$
& $-kC_{(\nu )\alpha }$ \\ \hline
\end{tabular}
$\vspace{1cm}
\end{center}

A metric space is $1+(n-1)$ decomposable if it can be written in the form:

\begin{equation}
ds^{2}=\epsilon (dx^{1})^{2}+g_{\alpha \beta }(x^{\gamma })dx^{\alpha
}dx^{\beta }  \label{sx2.6}
\end{equation}
where $\epsilon =\pm 1$, $g_{\alpha \beta }(x^{\gamma })$ is the metric in
the hypersurface $x^{1}=$const. and Greek indices take the values $2,3,...,n$%
. The following general result holds for the conformal algebra of these
spaces \cite{Tsamparlis-Nikolop-Apost}.

{\bf Proposition 1} \label{Prop2.1}{\em The proper CKVs of a }$1+(n-1)${\em %
\ metric are of the form: } 
\begin{equation}
X^{a}=f(x^{b})\delta _{1}^{a}+K^{\alpha }\delta _{\alpha }^{a}.
\label{sx2.7}
\end{equation}
{\em In this formula }$K^{\alpha }${\em \ is a proper CKV of the }$(n-1)$%
{\em \ metric given by: } 
\begin{equation}
K^{\alpha }=\frac{1}{p}m(x^{1})\phi ^{,\alpha }(x^{b})+L^{\alpha }(x^{j})
\label{sx2.8}
\end{equation}
{\em where:}

{\em 1. }$m(x^{1})${\em \ satisfies the equation:  }
\begin{equation}
m_{,11}+\epsilon pm=0.  \label{sx2.9}
\end{equation}

{\em 2. The vector }$\phi ^{,\alpha }(x^{j})${\em \ is a gradient CKV of the 
}$(n-1)${\em \ metric }$g_{\alpha \beta }${\em \ and satisfies the
condition: } 
\begin{equation}
\phi _{\mid \alpha \beta }(x^{c})=-p\phi g_{\alpha \beta }  \label{sx2.10}
\end{equation}
{\em where }$\mid ${\em \ denotes covariant differentiation wrt the }$(n-1)$%
{\em \ metric }$g_{\alpha \beta }.$

{\em 3. }$L^{\alpha }(x^{b})${\em \ is a non-gradient KV or HVF of the }$%
(n-1)-${\em metric and the }$f(x^{b})${\em \ is a }$C^{\infty }${\em \
function given by the formula: } 
\begin{equation}
f(x^{b})=-\frac{\epsilon }{p}m_{,1}\phi (x^{b}).  \label{sx2.11}
\end{equation}
{\em Furthermore the HVF }$H^{a}${\em \ of the }$n${\em \ metric (if it
exists) is given by: } 
\begin{equation}
H^{a}=bx^{1}\delta _{1}^{a}+L^{\alpha }\delta _{\alpha }^{a}  \label{sx2.12}
\end{equation}
{\em where }$L^{\alpha }${\em \ is the HVF (if it exists) vector of the }$%
(n-1)-${\em metric. The last equation implies that the KVs of the }$(n-1)-$%
{\em metric are identical with those of the }$1+(n-1)${\em \ decomposable
space.}

Using Proposition 1 and the previous considerations on the CKVs of
conformally related spaces we are able to compute the conformal algebra of a
metric conformal to a $1+(n-1)$ decomposable metric whose $(n-1)$ part is
the metric of a space of constant curvature. In particular for $n=4$ we
obtain the RW-like metrics which we study in the next section.

\section{The case of RW-like metrics}

The isotropic RW space-time admits a six dimensional symmetry group acting
on 3D spacelike orbits. According to the construction described in the
Introduction this symmetry group defines the generic metric:

\begin{equation}
ds^{2}=S^{2}(\tau )\left( \epsilon d\tau ^{2}+U^{2}(k,x^{\alpha })d\sigma
_{3}^{2}\right]  \label{sx3.1}
\end{equation}
where $\epsilon =\pm 1,U(k,x^{\alpha })=\left( 1+\frac{k}{4}x^{\alpha
}x_{\alpha }\right) ^{-1},$ $k=0,\pm 1$ and $d\sigma
_{3}^{2}=dx^{2}+dy^{2}+dz^{2}$. The generic metric (\ref{sx3.1}) we shall
call {\em RW-like metric.} The standard RW space-time is a member of the FOM
generated by the RW-like metric and the variable $\tau $ is related to the
standard variable $t$ (cosmic time in RW space-time) by the relation:

\begin{equation}
dt=S(\tau )d\tau .  \label{sx3.2}
\end{equation}
The RW-like metric is conformally related to Einstein space-time ($\epsilon
=-1,k=1$), the anti-Einstein space-time ($\epsilon =-1,k=-1$), the Minkowski
space-time ($\epsilon =-1,k=0$) and to a positive definite 1+3 decomposable
metric ($\epsilon =1$). In each case the 3-space is a space of constant
curvature $R=6k$.

By applying the results of the previous section for $n=4$ one computes
easily the complete conformal algebra of the metric (\ref{sx3.1}). The
results of the calculations are collected in Table\ II for the cases $k=\pm
1 $ and in Table\ III for the case $k=0.$ In both tables the quantities $%
\phi _{k}$, ${\bf H}$ and ${\bf C}_{\mu }$ are defined in Table I and for
convenience we have set $(c_{+},c_{-})\equiv ({\cosh \tau ,\cos \tau }$\ )
and $(s_{+},s_{-})\equiv \left( \sinh {\tau ,\sin \tau }\right) $.
Furthermore we give the forms of the ''scale factor'' $S(\tau )$ in order
the generic metric (\ref{sx3.1}) to admit extra KVs. We note that there are
either one or four extra KVs, a result which is in agreement with well known
theorems \cite{Hall-Steele}. \vspace{1cm}

{\small TABLE II. The nine {\em proper} CKVs of the RW-like metrics for the
case }$k=\pm 1${\small \ together with their conformal factors. The last
column gives }$S(\tau )${\small \ for extra KVs. }

\begin{center}
\begin{tabular}{|l|l|l|l|}
\hline
{\bf \#} & Proper {\bf CKV }${\bf X}$ & {\bf Conformal Factor} $\psi $ & 
\begin{tabular}{l}
$S(\tau )$ {\bf for} \\ 
{\bf extra KVs}
\end{tabular}
\\ \hline
{\it 1} & ${\bf Y}=\partial _{\tau }$ & $\left[ {\ln S(\tau )}\right]
_{,\tau }$ & $A$ \\ \hline
{\it 1} & ${\bf H}_{1}^{k}=\epsilon k\phi _{k}({\bf H})c_{\epsilon
k}\partial _{\tau }+{\bf H}s_{\epsilon k}$ & $\epsilon k\frac{\phi _{k}({\bf %
H})}{R(\tau )}\left[ S(\tau )c_{\epsilon k}\right] _{,\tau }$ & $%
A/c_{\epsilon k}$ \\ \hline
{\it 1} & ${\bf H}_{2}^{k}=\phi _{k}({\bf H})s_{\epsilon k}\partial _{\tau }+%
{\bf H}c_{\epsilon k}$ & $\frac{\phi _{k}({\bf H})}{R(\tau )}\left[ S(\tau
)s_{\epsilon k}\right] _{,\tau }$ & $A/s_{\epsilon k}$ \\ \hline
{\it 3} & ${\bf Q}_{\mu }^{k}=\epsilon k\phi _{k}({\bf C}_{\mu })c_{\epsilon
k}\partial _{\tau }+{\bf C}_{\mu }s_{\epsilon k}$ & $\epsilon k\frac{\phi
_{k}({\bf C}_{\mu })}{R(\tau )}\left[ S(\tau )c_{\epsilon k}\right] _{,\tau
} $ & $A/c_{\epsilon k}$ \\ \hline
{\it 3} & ${\bf Q}_{\mu +3}^{k}=\phi _{k}({\bf C}_{\mu })s_{\epsilon
k}\partial _{\tau }+{\bf C}_{\mu }c_{\epsilon k}$ & $\frac{\phi _{k}({\bf C}%
_{\mu })}{R(\tau )}\left[ S(\tau )s_{\epsilon k}\right] _{,\tau }$ & $%
A/s_{\epsilon k}$ \\ \hline
\end{tabular}
\vspace{1cm}
\end{center}

{\small TABLE III. The nine proper CKVs of the RW-like spaces (}$k=0${\small %
) together with their conformal factors. The last column shows }$S(\tau )$%
{\small \ for extra KVs (whenever they exist). }

\begin{center}
\begin{tabular}{|l|l|l|l|}
\hline
{\bf \#} & Proper {\bf CKV }${\bf X}$ & {\bf Conformal Factor} $\psi $ & 
\begin{tabular}{l}
$S(\tau )$ {\bf for} \\ 
{\bf extra KVs}
\end{tabular}
\\ \hline
{\it 1} & ${\bf P}_{{\tau }}=\partial _{{\tau }}$ & $(\ln S)_{,{\tau }}$ & $%
A $ \\ \hline
{\it 3} & ${\bf M}_{{\tau \alpha }}=x_{{\alpha }}\partial _{{\tau }%
}-\epsilon {\tau }\partial _{{\alpha }}$ & $x_{{\alpha }}(\ln S)_{,{\tau }}$
& $A$ \\ \hline
{\it 1} & ${\bf H}=x^{a}\partial _{a}$ & ${\bf H}(\ln S)+1$ & $1/A\tau $ \\ 
\hline
{\it 1} & ${\bf K}_{{\tau }}=2\epsilon {\tau }{\bf H}-\left(
x^{c}x_{c}\right) \partial _{{\tau }}$ & $-(\ln S)_{,\tau }\left( -\epsilon
\tau ^{2}+x^{2}+y^{2}+z^{2}\right) +2\epsilon \tau $ & $------$ \\ \hline
{\it 3} & ${\bf K}_{\mu }=2x_{\mu }{\bf H}-\left( x^{c}x_{c}\right) \partial
_{\mu a}$ & $2x_{\mu }\left[ \tau (\ln S)_{,\tau }+1\right] $ & $1/A\tau $
\\ \hline
\end{tabular}
\vspace{1cm}
\end{center}

In both cases the remaining eight or five vector fields are proper CKVs. If
we set $\epsilon =-1$ we find the CKVs of the standard RW metrics. These
vectors coincide with those found earlier \cite{Maartens-Maharaj}.

\section{Ricci Inheriting Collineations of RW space-times}

As it has been explained in the Introduction the proper RICs and the proper
RCs of a given space-time metric $g_{ab}$ are the extra CKVs and KVs
respectively of the generic metric $G_{ab}$ defined by the symmetry group of
the space-time metric $g_{ab}.$ In fact in this approach one essentially
sees the Ricci tensor $R_{ab}$ of $g_{ab}$ as a ''metric'' on the manifold $%
{\cal M}$. This metric is different from the space-time metric $g_{ab}$
because:

\begin{enumerate}
\item  It can be degenerate i.e. of $rank<4$.

\item  Its signature can change from point to point on the manifold.
\end{enumerate}

Due to the strong relation between the two metrics $g_{ab},R_{ab}$ in the
case of non-degeneracy, the results on the isometries (KVs) of $g_{ab}$
extend naturally to the RCs. For example the following result on RCs due to
Hall et al. \cite{Hall-Roy-Vaz} is a direct consequence of the corresponding
result on KVs of the original metric $g_{ab}$:

{\em If the Ricci tensor is of rank 4, at every point of the space-time
manifold, then the smooth (}$C^{2}$ {\em is enough) RCs form a Lie algebra
of smooth vector fields whose dimension is }$\leq 10${\em \ and }$\neq 9$%
{\em . This Lie algebra contains the proper RCs and their degeneracies. }

Concerning the RICs of the RW metric we expect that, due to the conformal
flatness of the RW-like metric, the maximum number of expected RICs is 15,
of which 6 are KVs and 9 proper RICs (one of which reduces to a KV or to
extra 4 KVs for certain forms of the conformal factor as we shall show below
in Table II and Table III). These vectors have already been found and are
given in the second column of Table II and Table III. What it remains to be
done is to compute the Ricci tensor and express their components in terms of
those of the Ricci tensor. {\em No further computations are needed!}

The Ricci tensor of the standard RW metric is:

\begin{equation}
R_{ab}=diag\left[ R_{0}(\tau ),R_{1}(\tau )U^{2}(k,x^{\alpha }),R_{1}(\tau
)U^{2}(k,x^{\alpha }),R_{1}(\tau )U^{2}(k,x^{\alpha })\right]  \label{sx4.1}
\end{equation}
where:

\begin{equation}
R_{0}(\tau )=\frac{3\left( \dot{S}^{2}-S\ddot{S}\right) }{S}  \label{sx4.2a}
\end{equation}
\begin{equation}
R_{1}(\tau )=\frac{S\ddot{S}+2\dot{S}^{2}+2kS^{2}}{S^{2}}  \label{sx4.2}
\end{equation}
and $U(k,x^{\alpha })=\left( 1+\frac{k}{4}x^{\alpha }x_{\alpha }\right)
^{-1} $.

We note immediately that this belongs to the family of metrics (\ref{sx3.1}%
), as expected. Because there is no guarantee that this tensor is
non-degenerate and of constant signature one has to consider the cases that
the Ricci tensor is degenerate and non-degenerate and in the latter case
consider further the two possible subcases of Euclidean and Lorentzian
signature.

\subsection{Ricci Inheritance Collineations of RW space-times in the
non-degenerate case}

In this case we write:

\begin{equation}
ds_{R}^{2}=R_{0}(\tau )d\tau ^{2}+R_{1}(\tau )U^{2}(k,x^{\mu })d\sigma
_{3}^{2}.  \label{sx4.3}
\end{equation}
Using the transformation ($R_{0}(\tau )R_{1}(\tau )\neq 0$):

\begin{equation}
d\overline{\tau }=\sqrt{\left| \frac{R_{0}(\tau )}{R_{1}(\tau )}\right| }%
d\tau \Rightarrow \partial _{\overline{\tau }}=\sqrt{\left| \frac{R_{1}(\tau
)}{R_{0}(\tau )}\right| }\partial _{\tau }  \label{sx4.4}
\end{equation}
we find the metric in the form:

\begin{equation}
ds_{R}^{2}=R_{1}(\overline{\tau })\left[ \epsilon d\overline{\tau }%
^{2}+U^{2}(k,x^{\alpha })d\sigma _{3}^{2}\right]  \label{sx4.5}
\end{equation}
where:

\begin{equation}
\epsilon =sign\left( \frac{R_{0}(\tau )}{R_{1}(\tau )}\right) .
\label{sx4.6}
\end{equation}
The metric (\ref{sx4.5}) is a non-degenerate metric of the form (\ref{sx3.1}%
) therefore its ($C^{\infty }$) RICs are the CKVs of the RW-like FOM (\ref
{sx3.1}). Furthermore the proper RCs of the RW metric are the extra KVs of
the generic metric (which occur for special forms of the ''conformal''
factor $R_{1}(\overline{\tau })$). Using the results of Table II and Table
III we write without any further computations all {\em proper} RICs by
changing $S(\tau )\leftrightarrow R_{1}(\overline{\tau })$ and $\tau
\leftrightarrow \overline{\tau }.$ The results for $k=\pm 1$ are listed in
Table\ IV and those for $k=0$ in Table\ V. The quantities $\phi _{k}$\ and
the vector fields ${\bf H},{\bf C}_{\mu }$\ of the 3-space of constant
curvature are defined in Table I. We have set again $(c_{+},c_{-})\equiv
\left[ \cosh \bar{\tau}(\tau ),\cos \bar{\tau}(\tau )\right] $\ , $%
(s_{+},s_{-})\equiv \left[ \sinh \bar{\tau}(\tau ),\sin \bar{\tau}(\tau
)\right] $. \pagebreak

{\small TABLE IV. The complete algebra of proper RICs of the RW space-times
for }$k=\pm 1${\small . }

\begin{center}
\begin{tabular}{|l|l|l|}
\hline
{\bf \#} & {\bf RICs X} & {\bf Conformal factor }${\bf \psi }$ \\ \hline
{\it 1} & ${\bf Y}=\left| \frac{R_{1}(\tau )}{R_{0}(\tau )}\right|
^{1/2}\partial _{\tau }$ & $\left| R_{1}(\tau )R_{0}(\tau )\right|
^{-1/2}\left( \left| R_{1}(\tau )\right| ^{1/2}\right) _{,\tau }$ \\ \hline
{\it 1} & ${\bf H}_{1}^{k}=\epsilon k\phi _{k}({\bf H})\left| \frac{%
R_{1}(\tau )}{R_{0}(\tau )}\right| ^{1/2}c_{\epsilon k}\partial _{\tau }+%
{\bf H}s_{\epsilon k}$ & $\epsilon k\phi _{k}({\bf H})\left| R_{1}(\tau
)R_{0}(\tau )\right| ^{-1/2}\left[ \left| R_{1}(\tau )\right|
^{1/2}c_{\epsilon k}\right] _{,\tau }$ \\ \hline
{\it 1} & ${\bf H}_{2}^{k}=\phi _{k}({\bf H})\left| \frac{R_{1}(\tau )}{%
R_{0}(\tau )}\right| ^{1/2}s_{\epsilon k}\partial _{\tau }+{\bf H}%
c_{\epsilon k}$ & $\phi _{k}({\bf H})\left| R_{1}(\tau )R_{0}(\tau )\right|
^{-1/2}\left[ \left| R_{1}(\tau )\right| ^{1/2}s_{\epsilon k}\right] _{,\tau
}$ \\ \hline
{\it 3} & ${\bf Q}_{\mu }^{k}=\epsilon k\phi _{k}({\bf C}_{\mu })\left| 
\frac{R_{1}(\tau )}{R_{0}(\tau )}\right| ^{1/2}c_{\epsilon k}\partial _{\tau
}+{\bf C}_{\mu }s_{\epsilon k}$ & $\epsilon k\phi _{k}({\bf C}_{\mu })\left|
R_{1}(\tau )R_{0}(\tau )\right| ^{-1/2}\left[ \left| R_{1}(\tau )\right|
^{1/2}c_{\epsilon k}\right] _{,\tau }$ \\ \hline
{\it 3} & ${\bf Q}_{\mu +3}^{k}=\phi _{k}({\bf C}_{\mu })\left| \frac{%
R_{1}(\tau )}{R_{0}(\tau )}\right| ^{1/2}s_{\epsilon k}\partial _{\tau }+%
{\bf C}_{\mu }c_{\epsilon k}$ & ${\phi }_{k}({\bf C}{_{\mu })}\left|
R_{1}(\tau )R_{0}(\tau )\right| ^{-1/2}\left[ \left| R_{1}(\tau )\right|
^{1/2}s_{\epsilon k}\right] _{,\tau }$ \\ \hline
\end{tabular}
\vspace{1cm}
\end{center}

{\small TABLE V. The complete algebra of proper RICs of the RW space-times
for }$k=0${\small . }

\begin{center}
\begin{tabular}{|l|l|l|}
\hline
{\bf \#} & {\bf RICs X} & {\bf Conformal factor }${\bf \psi }$ \\ \hline
{\it 1} & ${\bf P}_{\bar{\tau}}=\left| \frac{R_{1}(\tau )}{R_{0}(\tau )}%
\right| ^{1/2}\partial _{\tau }$ & $\psi ({\bf P}_{\bar{\tau}})=\frac{1}{2}%
\left| \frac{R_{1}(\tau )}{R_{0}(\tau )}\right| ^{1/2}(\ln R_{1})_{,\tau }$
\\ \hline
3 & ${\bf M}_{\mu \bar{\tau}}=-\epsilon \bar{\tau}(\tau )\partial _{\mu
}+x^{\mu }\left| \frac{R_{1}(\tau )}{R_{0}(\tau )}\right| ^{1/2}\partial
_{\tau }$ & $\psi ({\bf M}_{\alpha \tau })=\frac{1}{2}x^{\alpha }\left| 
\frac{R_{1}(\tau )}{R_{0}(\tau )}\right| ^{1/2}(\ln R_{1})_{,\tau }$ \\ 
\hline
{\it 1} & ${\bf H}=\bar{\tau}(\tau )\left| \frac{R_{1}(\tau )}{R_{0}(\tau )}%
\right| ^{1/2}\partial _{\tau }+x^{\mu }\partial _{\mu }$ & $\psi ({\bf H})=%
\frac{1}{2}\bar{\tau}(\tau )\left| \frac{R_{1}(\tau )}{R_{0}(\tau )}\right|
^{1/2}(\ln R_{1})_{,\tau }+1$ \\ \hline
4 & 
\begin{tabular}{l}
${\bf K}_{a}=\left[ 2\epsilon \delta _{\tau a}\bar{\tau}(\tau )+2x^{\mu
}\delta _{\mu a}\right] {\bf H}-$ \\ 
$-\left[ \epsilon \bar{\tau}^{2}(\tau )+x^{\beta }x_{\beta }\right] \partial
_{a}$%
\end{tabular}
& $
\begin{array}{l}
\psi ({\bf K}_{\tau })=-\frac{1}{2}\left| \frac{R_{1}(\tau )}{R_{0}(\tau )}%
\right| ^{1/2}(\ln R_{1})_{,\tau } \\ 
\times \left[ \bar{\tau}^{2}(\tau )+x^{2}+y^{2}+z^{2}\right] -2\bar{\tau}%
(\tau ) \\ 
\psi ({\bf K}_{\mu })=2x^{\mu }\left[ \frac{1}{2}\bar{\tau}(\tau )\left| 
\frac{R_{1}(\tau )}{R_{0}(\tau )}\right| ^{1/2}(\ln R_{1})_{,\tau }+1\right]
\end{array}
$ \\ \hline
\end{tabular}
\end{center}

To demonstrate how these results are obtained from those of Table\ II and
Table\ III let us consider the case of the RIC ${\bf P}_{\bar{\tau}}$. This
corresponds to the CKV ${\bf P}_{\tau }$ of the RW-like metrics (\ref{sx3.1}%
) whose conformal factor is $\psi ({\bf P}_{\tau })=\left[ \ln S(\tau
)\right] _{,\tau }$. To compute ${\bf P}_{\bar{\tau}},\psi ({\bf P}_{\bar{%
\tau}})$ we make the correspondence $S(\tau )\leftrightarrow R_{1}(\overline{%
\tau })$ and $\tau \leftrightarrow \overline{\tau }$ in (\ref{sx4.4}) and
write down the result immediately. For the other vectors we work in a
similar manner being careful to keep the unknown function $\bar{\tau}(\tau )$
wherever it occurs.

\subsection{Ricci inheritance collineations of the RW space-times in the
degenerate case}

In order to obtain results comparable with the ones in the existing
literature we consider the RW space-time metric in spherical coordinates: 
\[
ds^{2}=-dt^{2}+S^{2}(t)\left( \frac{dr^{2}}{1-kr^{2}}+r^{2}d\theta
^{2}+r^{2}\sin ^{2}\theta d\phi ^{2}\right) . 
\]
The Ricci tensor is again diagonal with elements: 
\begin{eqnarray}
R_{00} &=&-3S_{,tt}/S  \label{sx4.7} \\
R_{\alpha \beta } &=&\frac{\Delta }{S^{2}}g_{\alpha \beta }  \label{sx4.8}
\end{eqnarray}
where $g_{\alpha \beta }$ is the metric of the 3-space of constant curvature
and the quantity: 
\begin{equation}
\Delta =S\ddot{S}+2\dot{S}^{2}+2k.  \label{sx4.9}
\end{equation}
The condition for the Ricci tensor to be degenerate is $\det R_{ab}=0.$ This
implies that there are two cases to consider $R_{00}=0$, $R_{\alpha \beta
}\neq 0$ ($\Delta \neq 0$) and $R_{00}\neq 0$, $R_{\alpha \beta }=0$ ($%
\Delta =0$). We have the following result.

{\bf Proposition 2} {\em The RW space-times with a degenerate Ricci tensor
admit infinite RICs as follows:}

{\em (a) Case }$R_{00}=0${\em , }$\Delta \neq 0.$

{\em The scale factor is }$S(t)=at+b${\em \ where }$a,b${\em \ are constants
and the RICs are of the form } 
\begin{equation}
X=f(x^{a})\partial _{t}+X_{\perp }^{\mu }(x^{\mu })\partial _{\mu }
\label{sx4.10}
\end{equation}

{\em where }$X_{\perp }^{\mu }(x^{\mu })${\em \ is a CKV of the 3-metric }$%
g_{\alpha \beta }${\em \ and }$f(x^{a})${\em \ is an arbitrary smooth
function.}

{\em (b) Case }$R_{00}\neq 0${\em , }$\Delta =0.$

{\em The form of the RICs is:}

\begin{equation}
X=\frac{h(t)}{\left| R_{00}\right| ^{1/2}}\partial _{t}+X^{\mu
}(x^{a})\partial _{\mu }  \label{sx4.10c}
\end{equation}

{\em where }$X^{\mu }(x^{a})${\em \ are arbitrary (but smooth) functions of
the space-time coordinates.}

{\bf Proof}

We write the Ricci Inheritance equations ${\cal L}_{{\bf \xi }}R_{ab}=2\psi
R_{ab}$ in detailed form:

\begin{eqnarray}
R_{00,0}X^{0}+2R_{00}X_{,0}^{0} &=&2\psi R_{00}  \label{sx4.10d} \\
R_{00}X_{,\mu }^{0}+R_{\mu \mu }X_{,0}^{\mu } &=&0  \label{sx4.10e} \\
R_{\mu \mu ,0}X^{0}+\sum_{\nu }R_{\mu \mu ,\nu }X^{\nu }+2R_{\mu \mu
}X_{,\mu }^{\mu } &=&2\psi R_{\mu \mu }  \label{sx4.10f} \\
R_{\mu \mu }X_{,\nu }^{\mu }+R_{\nu \nu }X_{,\mu }^{\nu } &=&0%
\mbox{ (no sum
over }\mu ,\nu )  \label{sx4.10g}
\end{eqnarray}

(a) Case $R_{00}=0$, $\Delta \neq 0.$

Equation (\ref{sx4.10d}) implies $X^{0}=f(x^{a})$ where $f(x^{a})$ is an
arbitrary (smooth) function of the co-ordinates. Furthermore eq. (\ref
{sx4.10e}) gives $X^{\mu }=X^{\mu }(x^{\mu })$ which together with $%
S(t)=at+b $ means that $\Delta =const.$ Hence the space-time is $1+3$
decomposable. Using eq. (\ref{sx3.1}) it easy to show that equations (\ref
{sx4.10f}) and (\ref{sx4.10g}) can be written in concise form as: 
\begin{equation}
{\cal L}_{{\bf X}_{\perp }}g_{\mu \nu }=2\psi (x^{\rho })g_{\mu \nu }
\label{sx3.14}
\end{equation}
where ${\bf X}_{\perp }=X^{\mu }(x^{\mu })\partial _{\mu }$ is the spatial
part of the symmetry vector and $g_{\mu \nu }$ is the 3-metric of the
hypersurfaces $t=const.$ of constant curvature: 
\begin{equation}
g_{\mu \nu }=U^{2}(x^{\mu })\delta _{\mu \nu }.  \label{sx3.15}
\end{equation}
The last equation implies that ${\bf X}_{\perp }$ is a CKV\ of the 3-metric $%
g_{\mu \nu }$ and can be found from the results of Section II (Table I)
setting $n=3$. (This result agrees with the general result of Hall et al. on 
$1+3$ decomposable space-times \cite{Hall-Roy-Vaz}). However due to the fact
that the component $X^{0}$ of the RIC is arbitrary the Lie algebra is
infinite dimensional.

(b) $R_{00}\neq 0$, $\Delta =0.$

Equations (\ref{sx4.10d}), (\ref{sx4.10e}) imply that the quantities $%
X^{0}(t)$ and $\psi (t)$ depend only on the timelike coordinate $t$. We
introduce the positive function $h(t)$ as follows:

\[
\psi (t)=2X^{0}(t)\left[ \ln h(t)\right] _{t} 
\]
Then (\ref{sx4.10d}) gives $X^{0}=\frac{h(t)}{\sqrt{\left| R_{00}\right| }}$%
. The rest of the proof follows easily.

For degenerate Ricci tensor, RICs are infinite therefore mathematically they
are of no interest. From the physical point of view they do lead to
restrictions on the scale factor and consequently on the matter fluid. Thus
it has been shown \cite{Nunez-Percoco-Villalba} that in the case $R_{00}=0$
the equation of state is $\rho +3p=0$ whereas for $\Delta =0$ the equation
of state is $\rho =p$ (stiff matter). However both equations of state are of
limiting physical interest.

\section{Ricci collineations of the RW space-time}

We consider the non-degenerate case because in the degenerate case we have
infinite RCs which is of no interest. The proper RCs of the RW space-time
are obtained form Table II and Table III by setting the conformal factor of
each vector equal to zero. The results are given in the last column of Table
IV and Table V from which it follows that there are two cases to be
considered: $R_{1}(\tau )=const$ and $R_{1}(\tau )\neq const.$ The results
of the computations for the cases $k=\pm 1$ are collected in Table VI and
those of $k=0$ in Table VII. We have introduced for convenience the new time
coordinate $\tilde{\tau}$ by the relation:

\begin{equation}
\tilde{\tau}(\tau )=\int \left| R_{0}(\tau )\right| ^{1/2}d\tau .
\label{sx4.30}
\end{equation}
\vspace{1cm}

{\small TABLE VI. The proper RCs of the RW space-times for }$k=\pm 1${\small %
\ and the expression of }$R_{1}$ {\small for which the corresponding
collineations are admitted. \ A is an integration constant. }

\begin{center}
\begin{tabular}{|l|l|l|}
\hline
{\bf \#} & {\bf RCs X (}$k=\pm 1)$ & ${\bf R}_{1}(\tau )$ \\ \hline
{\it 1} & ${\bf Y}=A\partial _{\tilde{\tau}}$ & $A$ \\ \hline
{\it 1} & ${\bf H}_{1}^{k}=\epsilon k\phi _{k}({\bf H})A\partial _{\tilde{%
\tau}}+{\bf H}ta_{-\epsilon k}(\frac{\tilde{\tau}}{A})$ & $A^{2}c_{-\epsilon
k}^{2}(\frac{\tilde{\tau}(\tau )}{A})$ \\ \hline
{\it 1} & ${\bf H}_{2}^{\epsilon }=\phi _{\epsilon }({\bf H})A\partial _{%
\tilde{\tau}}-{\bf H}\coth (\frac{\tilde{\tau}}{A})$ & $A^{2}\sinh ^{2}(%
\frac{\tilde{\tau}(\tau )}{A})$ \\ \hline
{\it 3} & ${\bf Q}_{\mu }^{k}=\epsilon k\phi _{k}({\bf C}_{\mu })A\partial _{%
\tilde{\tau}}+{\bf C}_{\mu }ta_{-\epsilon k}(\frac{\tilde{\tau}}{A})$ & $%
A^{2}c_{-\epsilon k}^{2}(\frac{\tilde{\tau}(\tau )}{A})$ \\ \hline
{\it 3} & ${\bf Q}_{\mu +3}^{k}=\phi _{k}({\bf C}_{\mu })A\partial _{\tilde{%
\tau}}-{\bf C}_{\mu }\coth (\frac{\tilde{\tau}}{A})$ & $A_{k}^{2}\sinh ^{2}(%
\frac{\tilde{\tau}(\tau )}{A})$ \\ \hline
\end{tabular}
\pagebreak
\end{center}

{\small TABLE VII. The proper RCs of the RW space-times for }$k=0.${\small \
A is an integration constant. }

\begin{center}
\begin{tabular}{|l|l|l|}
\hline
{\bf \#} & {\bf RCs X (}$k=0)$ & $R_{1}(\tau )$ \\ \hline
{\it 1} & ${\bf P}_{\tilde{\tau}}=\left| A\right| ^{1/2}\partial _{\tilde{%
\tau}}$ & $A$ \\ \hline
{\it 3} & ${\bf M}_{\alpha \tilde{\tau}}=x^{\alpha }\left| A\right|
^{1/2}\partial _{\tilde{\tau}}-\epsilon \tilde{\tau}(\tau )\left| A\right|
^{-1/2}\partial _{\alpha }$ & $A$ \\ \hline
{\it 1} & ${\bf H}=A\partial _{\tilde{\tau}}+x^{\alpha }\partial _{\alpha }$
& $\epsilon A^{2}e^{-2\tilde{\tau}(\tau )/A}$ \\ \hline
{\it 3} & ${\bf K}_{\alpha }=2x_{\alpha }{\bf H}-\left( \epsilon e^{2\tilde{%
\tau}/A}+x^{\beta }x_{\beta }\right) \partial _{\alpha }$ & $\epsilon
A^{2}e^{-2\tilde{\tau}(\tau )/A}$ \\ \hline
\end{tabular}
\vspace{1cm}
\end{center}

From Table VI and Table VII we have the following result:

{\bf Proposition 3} {\em RW space-times with }$k=\pm 1${\em \ admit exactly
one proper RC when the spatial component }$R_{1}=A\neq 0${\em \ and four
when }$R_{1}${\em \ is of the form }$A^{2}\sinh ^{2}(\frac{\tilde{\tau}(\tau
)}{A})${\em \ or }$A^{2}c_{-\epsilon k}^{2}(\frac{\tilde{\tau}(\tau )}{A})$%
{\em . \label{Prop4.4}The flat RW space-times (}$k=0${\em ) admit four
proper RCs when the spatial component }$R_{1}${\em \ of the Ricci tensor
equals }$R_{1}=A\neq 0${\em \ and }$R_{1}=\epsilon A^{2}e^{-2\tilde{\tau}%
(\tau )/A}${\em . In both cases }$A${\em \ is a real constant and }$\tilde{%
\tau}(\tau )${\em \ is defined in (\ref{sx4.30}).}

\section{Matter inheritance collineations of the RW space-times\label{MICs}%
\label{MICCs}}

MICs and MCs of the RW metric are calculated in the same manner as the RICs
and RCs. Replacing $R_{ab}$ from (\ref{sx4.2}), (\ref{sx4.2a}) in Einstein
field equations:

\begin{equation}
T_{ab}=R_{ab}-\frac{1}{2}Rg_{ab}  \label{sx5.1a}
\end{equation}
and the generic metric (\ref{sx3.1}) for $\epsilon =-1$ we compute: 
\[
T_{ab}=diag\left[ T_{0}(\tau ),T_{1}(\tau )U^{2}(x^{\mu }),T_{1}(\tau
)U^{2}(x^{\mu }),T_{1}(\tau )U^{2}(x^{\mu })\right] 
\]
where: 
\begin{equation}
T_{0}=\frac{3(S,_{\tau })^{2}}{S^{2}}\qquad ,\qquad T_{1}=\frac{-2SS,_{\tau
\tau }+(S,_{\tau })^{2}}{S^{2}}  \label{sx5.1b}
\end{equation}
are functions of the time coordinate $\tau $. As expected this is also a
RW-like metric. In the case of non-degeneracy $T_{0}(\tau )T_{1}(\tau )\neq
0 $ we write:

\begin{equation}
T_{ab}=T_{1}(\tau )diag\left[ \frac{T_{0}(\tau )}{T_{1}(\tau )},U^{2}(x^{\mu
}),U^{2}(x^{\mu }),U^{2}(x^{\mu })\right] .  \label{sx5.1}
\end{equation}
Therefore the MICs and MCs can be written down immediately from the results
of Table II and Table III, if one makes the replacements $S(\tau
)\leftrightarrow T_{1}(\overline{\tau })$ and $\tau \leftrightarrow 
\overline{\tau }$ where $\bar{\tau}(\tau )$ is defined as follows:

\begin{equation}
d\bar{\tau}=\sqrt{\left| \frac{T_{0}(\tau )}{T_{1}(\tau )}\right| }d\tau
\Rightarrow \partial _{\overline{\tau }}=\sqrt{\left| \frac{T_{1}(\tau )}{%
T_{0}(\tau )}\right| }\partial _{\tau }.  \label{sx5.2}
\end{equation}
We note that the results are similar with those of RICs and RCs collected in
tables IV, V and VI, VII respectively provided that one makes the
replacements $R_{0}(\tau )\leftrightarrow T_{0}(\tau )$, $R_{1}(\tau
)\leftrightarrow T_{1}(\tau )$. Obviously there is no need to write
explicitly these tables again. Furthermore Proposition 3 applies equally
well to MICs and MCs.

\section{Applications}

Consider the standard RW cosmological model with vanishing cosmological
constant and comoving observers $u^{a}=S^{-1}(\tau )\delta _{\tau }^{a}$,
where $\tau =\int \frac{dt}{S(t)}$ and $t$ being the standard cosmic time.
For these observers the energy momentum tensor has a perfect fluid form i.e. 
$T_{ab}=\mu u_{a}u_{b}+ph_{ab}$ where $\mu ,p$ are the energy density and
the isotropic pressure measured by the observers $u^{a}$. From this
decomposition of $T_{ab}$ and in the coordinates $(\tau ,x^{\mu })$ one has:

\begin{equation}
T_{00}=\mu S^{2}(\tau ),T_{11}=pS^{2}(\tau )U^{2}(k,x^{\mu }).  \label{sx6.1}
\end{equation}
On the other hand using Einstein equations one computes the components $%
T_{00}(S,S_{,\tau },S_{,\tau \tau },U,)$ and $T_{11}(S,S_{,\tau },S_{,\tau
\tau },U,)$. These two results allow us to compute $\mu ,p$ in terms of $k$
and $S(\tau )$. Therefore the field equations leave one variable (the $%
S(\tau )$) free and one has to supplement an extra condition to solve the
model. This extra equation is a barotropic equation of state $p=p(\mu ).$
The obvious choice is a linear equation of state $p=(\gamma -1)\mu $. There
are several solutions for this simple choice which are of cosmological
interest.\ For example $\gamma =1$ $(p=0)$ implies degeneracy of the energy
momentum-tensor (dust) and the value $\gamma =\frac{4}{3}$ ($p=\frac{1}{3}%
\mu $) implies radiation dominated matter. Both states of matter are extreme
and they have been relevant at certain stages of the evolution of the
Universe. For other values of $\gamma $ one obtains intermediate states
which cannot be excluded as unphysical (see e.g. \cite{Kramer} for a
thorough review). However it appears that there does exist some uncertainty
in the choice on the value of $\gamma $ and of course a bigger one in the
choice of a linear equation of state. That is, one would be interested to
use a non-linear equation of state and deal with more complex forms of
matter, but there does not seem to exist an ''objective'' criterion for
writing down such an equation.

One such criterion can be established by the ''higher'' symmetries we
considered in the previous sections. Indeed one can look upon the RICs/RCs
and MICs/MCs as ''aesthetic'' symmetries which can provide via the equation
defining them, an extra equation generating an equation of state and
consequently a definite RW cosmological model. The reason for the use of
these symmetries is twofold:

a. They solve the model completely, that is, they allow us to compute all
kinematic and dynamic quantities involved in the RW cosmology

b. As we have shown, they do not violate either the geometry or the coupling
of the geometry to Physics.

In the following we shall follow this point of view and we shall use MCs to
define the equation of state, that is, we shall determine {\em all} flat ($%
k=0$) RW cosmological models which admit a MC. The reason for using MCs is
that they are directly related to the components of the energy momentum
tensor and one expects that they will have immediate and stronger physical
implications.

Writing Table VII\ in terms of the energy momentum tensor as explained in
section \ref{MICCs} we see that there are two cases to consider i.e. $%
T_{1}=T_{11}=A=const$ and $T_{1}=\varepsilon A^{2}e^{-2\tilde{\tau}(\tau )A}$
where the ''time'' coordinate $\tilde{\tau}$ is defined as follows: 
\begin{equation}
\tilde{\tau}(\tau )=\int \left| T_{0}(\tau )\right| ^{1/2}d\tau =\int \sqrt{%
\mu }Sd\tau  \label{sx6.2a}
\end{equation}
and we have used equations (\ref{sx3.2}) and (\ref{sx6.1}).

From eqs. (\ref{sx6.1}) and (\ref{sx5.1b}) we obtain:

\begin{equation}
\mu =\frac{3(S,_{\tau })^{2}}{S^{4}},\mbox{ }p=\frac{-2SS,_{\tau \tau
}+(S,_{\tau })^{2}}{S^{4}}  \label{sx6.4}
\end{equation}
which express the dynamic variables $\mu ,p$ in terms of the scale factor $%
S(\tau )$. Two other important kinematic quantities in the RW universe are
the Hubble ''constant'' $H$ and the deceleration parameter $q$ defined as
follows:

\begin{equation}
H=\frac{1}{3}\theta =\frac{S,_{\tau }}{S^{2}}\qquad ,\qquad q=1-\frac{%
SS,_{\tau \tau }}{(S,_{\tau })^{2}}.  \label{sx6.4a}
\end{equation}

\underline{Case I: $T_{1}(\tau )=A\equiv \varepsilon _{1}a^{2}$ ($%
\varepsilon _{1}=\pm 1,a\in R)$}

The constraint $T_{1}(\tau )=\varepsilon _{1}a^{2}$ leads to the condition:

\begin{equation}
pS^{2}(\tau )=\varepsilon _{1}a^{2}  \label{sx6.5}
\end{equation}
which by means of the second of (\ref{sx6.4}) gives the equation:

\begin{equation}
-2SS,_{\tau \tau }+(S,_{\tau })^{2}=\varepsilon _{1}a^{2}S^{2}.
\label{sx6.6}
\end{equation}
The solution of the differential equation (\ref{sx6.6}) provides the unknown
scale factor $S(\tau )$ and describes the RW model completely. To solve
equation (\ref{sx6.6}) we write it in the form:

\begin{equation}
2\left( \frac{S,_{\tau }}{S}\right) ,_{\tau }+\left( \frac{S,_{\tau }}{S}%
\right) ^{2}=-\varepsilon _{1}a^{2}  \label{sx6.7}
\end{equation}
which can be integrated easily. In Table VIII we present all four solutions
of (\ref{sx6.7}) together with the physical variables of the model that is,
energy density ($\mu $), isotropic pressure ($p$)$,$ Hubble constant ($H$)
and deceleration parameter ($q$). \vspace{1cm}

{\small TABLE VIII.\ The }$k=0${\small \ RW cosmological models which admit
the MCs }$P_{\tilde{\tau}},M_{\mu \tilde{\tau}}$.

\begin{center}
\begin{tabular}{|l|l|l|l|l|l|l|}
\hline
{\bf Case} & $S(\tau )$ & $\mu (\tau )$ & $p(\tau )$ & $H(\tau )$ & $q(\tau
) $ & {\bf Restrictions} \\ \hline
1 & $Be^{a\tau }$ & $-\frac{3A}{B^{2}e^{2a\tau }}$ & $\frac{A}{%
B^{2}e^{2a\tau }}$ & $\frac{3a}{Be^{a\tau }}$ & $0$ & $\varepsilon
_{1}=-1,A<0$ \\ \hline
2 & $B\cos ^{2}\frac{a\tau }{2}$ & $\frac{12A\left( 1-\cos a\tau \right) }{%
B^{2}\left( 1+\cos a\tau \right) ^{3}}$ & $\frac{4A}{B^{2}\left( 1+\cos
a\tau \right) ^{2}}$ & $\frac{2a\sin a\tau }{B\left( 1+\cos a\tau \right)
^{2}}$ & $\frac{1+\cos a\tau }{\sin ^{2}a\tau }$ & $\varepsilon _{1}=1,A>0$
\\ \hline
3 & $B\sinh ^{2}\frac{a\tau }{2}$ & $-\frac{3A}{B^{2}}\frac{\coth ^{2}\frac{%
a\tau }{2}}{\sinh ^{4}\frac{a\tau }{2}}$ & $\frac{A}{B^{2}}\sinh ^{-4}\frac{%
a\tau }{2}$ & $\frac{a}{B}\frac{\coth \frac{a\tau }{2}}{\sinh ^{2}\frac{%
a\tau }{2}}$ & $\frac{1}{2\cosh ^{2}a\tau }$ & $
\begin{array}{l}
\varepsilon _{1}=-1,A<0 \\ 
\left( \frac{S,_{\tau }}{S}\right) ^{2}>a^{2}
\end{array}
$ \\ \hline
4 & $B\cosh ^{2}\frac{a\tau }{2}$ & $-\frac{3A}{B^{2}}\frac{\tanh ^{2}\frac{%
a\tau }{2}}{\cosh ^{4}\frac{a\tau }{2}}$ & $\frac{A}{B^{2}}\cosh ^{-4}\frac{%
a\tau }{2}$ & $\frac{a}{B}\frac{\tanh \frac{a\tau }{2}}{\cosh ^{2}\frac{%
a\tau }{2}}$ & $-\frac{1}{2\sinh ^{2}a\tau }$ & $\left( \frac{S,_{\tau }}{S}%
\right) ^{2}<a^{2}$ \\ \hline
\end{tabular}
\end{center}

It is a straightforward matter (e.g. by using any algebraic computing
program) to check that indeed all four solutions of RW\ space-times of
Table\ VIII admit the MCs given in Table VII. A detailed study shows that
all MCs are proper, except the $P_{\tilde{\tau}}$ for case 1 which
degenerates to a HVF. Furthermore {\em all the energy conditions} are
satisfied.

Concerning the determination of the equation of state we use the energy
conservation equation (i.e. $T_{;a}^{\tau a}=0$) which in the coordinates $%
(\tau ,x^{\mu })$ gives:

\begin{equation}
\mu _{,\tau }=-3H(\mu +p)S.  \label{sx6.9}
\end{equation}
From the symmetry condition (\ref{sx6.5}) we compute:

\begin{equation}
2HSp=-p_{,\tau }.  \label{sx6.10}
\end{equation}
Eliminating $S(\tau )$ from the last two relations we find:

\begin{equation}
\frac{dp}{d\mu }=\frac{p_{,\tau }}{\mu _{,\tau }}=\frac{2}{3}\frac{p}{p+\mu }%
.  \label{sx6.11}
\end{equation}
This equation has two solutions:

\begin{equation}
p=-\frac{1}{3}\mu  \label{sx6.12}
\end{equation}
and:

\begin{equation}
\mu -\frac{3B}{a}\mid p\mid ^{3/2}+3p=0.  \label{sx6.13}
\end{equation}
For $B=0$ we obtain the first solution, which is a linear equation of state
with $\gamma =\frac{2}{3}$. It corresponds to the solution of case 1 of
Table VIII whose metric is:

\begin{equation}
ds^{2}=-dt^{2}+t^{\frac{4}{3\gamma }}(dx^{2}+dy^{2}+dz^{2}).  \label{sx6.7a}
\end{equation}
This space-time admits a HVF \cite{Wainright-Ellis} represented by the
vector $P_{\tilde{\tau}}$. The rest three vector fields are {\em proper MCs}.

The other solution of (\ref{sx6.11}) ($B\neq 0$) leads to a non-linear
equation of state and concerns the cases 2,3,4 of Table\ VIII.

\underline{Case II: $T_{1}(\tau )=\varepsilon A^{2}e^{-2\tilde{\tau}(\tau
)A} $ ($A\equiv \varepsilon _{1}a^{2}$ ,$\varepsilon _{1}=\pm 1,a\in R)$}

From eqs (\ref{sx6.2a}) and (\ref{sx6.4}) we compute $\tilde{\tau}$ in terms
of the scale factor:

\begin{equation}
\tilde{\tau}(\tau )=\int \sqrt{3}\left[ \ln S(\tau )\right] _{,\tau }d\tau
=\ln [S^{\sqrt{3}}].  \label{sx6.14}
\end{equation}
Using the last equation and the second of (\ref{sx6.1}) ($k=0$) we obtain:

\begin{equation}
p=p_{0}S^{-3B}  \label{sx6.15}
\end{equation}
where $B=\frac{2}{3}\left( 1+\frac{\sqrt{3}}{A}\right) $ and $%
p_{0}=\varepsilon A^{2}$. Furthermore eqs (\ref{sx6.4}) and (\ref{sx6.15})
give:

\begin{equation}
2SS,_{\tau \tau }-(S,_{\tau })^{2}=-p_{0}S^{4-3B}  \label{sx6.16}
\end{equation}
whose solution is not simple to find.

However we can find the equation of state. From (\ref{sx6.15}) we compute: 
\begin{equation}
p_{,\tau }=-3BpSH  \label{sx6.17}
\end{equation}
which, when combined with (\ref{sx6.9}), gives the following equation among
the dynamic variables $\mu ,p$: 
\begin{equation}
\frac{dp}{d\mu }=\frac{Bp}{\mu +p}.  \label{sx6.18}
\end{equation}
We consider two subcases.

\underline{$B=1\Leftrightarrow A=2\sqrt{3}$}

The solution of (\ref{sx6.18}) is:

\begin{equation}
\mu =p\mid \ln Cp\mid ,C=const,Cp>0  \label{sx6.19}
\end{equation}
which is always a non-linear equation.

\underline{$B\neq 1\Leftrightarrow A\neq 2\sqrt{3}$}

In this case the solution of (\ref{sx6.18}) is:

\begin{equation}
p-Dp^{1/3}=(B-1)\mu ,\mbox{ }D=const.,B\neq 0,1.  \label{sx6.20}
\end{equation}
Note that $D\neq 0$ and we have always a non-linear barotropic equation of
state.

From the above we conclude that:

{\bf Proposition 4} {\em The only perfect fluid and flat RW universe with a
linear equation of state which admits proper MCs is the RW model (\ref
{sx6.7a}) for }$\gamma =\frac{2}{3}${\em .}

\section{Discussion}

Nearly all known solutions of General Relativity involve some symmetry
assumption for the space-time metric. The symmetry is propagated in an
intrinsic manner to the higher levels of geometry (by covariant
differentiation) and to the Physics (by Einstein field equations). These
facts are in complete agreement with the fundamental approach of General
Relativity and perhaps they have also helped Einstein in the development of
the theory. The most important quantities effected by the symmetry
assumption are the Ricci tensor and the energy momentum tensor.

One usually treats the Ricci tensor and the energy momentum tensor as
independent of the metric without making use of the fact that these tensors
admit the isometry group $G$ of the metric, therefore they all are members
of the same family of second order tensors defined by the group $G$. As we
have shown it is the Lorentzian signature and the non-degeneracy which
distinguish the metric from the other two tensors. A further differentiation
between them is done by the energy conditions and the direct kinematic and
dynamic interpretation of the energy momentum tensor by a congruence of
observers.

One interesting question is if there are metrics whose Ricci or Matter
tensor is both non-degenerate and has Lorentzian character, therefore it can
be used equally well as a space-time metric with the same symmetries as the
original space-time metric! This is indeed possible. For example the RW
metric:

\begin{equation}
ds^{2}=-dt^{2}+\cosh ^{2}t\left( dx^{2}+dy^{2}+dz^{2}\right) .  \label{sx7.1}
\end{equation}
has the Ricci tensor: 
\begin{equation}
R_{00}=-3\qquad ,\qquad R_{\alpha \beta }=\frac{3}{2}\cosh 2t-\frac{1}{2}
\label{sx7.2}
\end{equation}
which is again a RW metric.

The results obtained in this paper can be summarized as follows.

a. The determination of the proper symmetries of $R_{ab},T_{ab}$ does not
require any new calculations because they follow as the extra symmetries of
the generic metric (if they exist) which is defined by the KVs of the
space-time metric. Indeed in Sections V and VI we computed all RCs and MCs
collineations of RW space-times using the KVs of the generic RW like metric (%
\ref{sx3.1}).

b. The collineation tree introduced by Katzin et al. should be reconsidered.
That is (for space-time metrics with symmetry) the essential part of the
tree ends in the Projective collineations, the rest of them being subcases
(except perhaps very few peculiar or uninteresting collineations) which
follow as special cases of the basic collineations.

c. RCs, MCs etc. are useful in supplying external constraints to the field
equations in order to make the system of field equations autonomous. Because
the equations resulting from these constraints are compatible with the
assumed symmetry of space-time and the restrictions imposed on Geometry and
Physics one expects that they can lead to physically viable models. Indeed
in Section VII we have shown that the geometric assumption that the flat RW
space-time admits a MC leads to a concrete non-linear (in general) equation
of state and consequently to definite RW cosmological models which satisfy
the energy conditions and are physically acceptable.

d. The energy momentum tensor attains a physical interpretation in terms of
energy density, pressure etc. only ${\em after}$ a congruence of observers
has been selected. For example, if one considers tilted observers \cite
{Ellis-Matravers-Treciokas,Coley-Review} one obtains a RW cosmological model
with a non-perfect fluid. This means that equations of state imposed on the
dynamical quantities $\mu ,p$ etc. hold for this concrete congruence of
observers and do not hold necessarily for other observers. However equations
of state defined by means of a symmetry assumption (e.g. a MC) are
independent of the observers and apply to the space-time Physics universally.

{\bf Acknowledgments}

We wish to thank Prof. MacCallum for useful correspondence.

\end{document}